\documentclass[twocolumn,amsmath,amssymb,amsfonts,superscriptaddress,floatfix,showpacs]{revtex4-1}
\pdfoutput=1
\usepackage{graphicx}% Include figure files
\usepackage{dcolumn}% Align table columns on decimal point
\usepackage{keyval}
\usepackage{bm}% bold math
\usepackage{subfigure}
\usepackage{float}
\usepackage{amsmath}
\usepackage[usenames,dvipsnames]{color}
\usepackage[normalem]{ulem}
\usepackage{array} 
\usepackage{physics}

\begin{document}

\title{
Inhomogeneous magnon scattering during ultrafast demagnetization
}
%Electronic structure and magnetic interactions in (Ni$_{0.8}$Fe$_{0.2}$)$_{1-x}$Cu$_x$ alloys with implications for ultrafast demagnetization}
%\title{Localization of Fe d-states in Ni-Fe-Cu alloys and implications for ultrafast demagnetization}

\author{Ronny Knut} \affiliation{Quantum Electromagnetics Division, National Institute of Standards and Technology, Boulder, CO 80305, USA}
\affiliation{Department of Physics and JILA, University of Colorado and NIST, Boulder, CO 80309, USA}
 \affiliation{Department of Physics and Astronomy, Uppsala University, Box 516, 75120 Uppsala, Sweden}

\author{Erna K. Delczeg-Czirjak} 
 \affiliation{Department of Physics and Astronomy, Uppsala University, Box 516, 75120 Uppsala, Sweden}

  \author{Somnath Jana} 
 \affiliation{Department of Physics and Astronomy, Uppsala University, Box 516, 75120 Uppsala, Sweden}

\author{Justin M. Shaw} 
 \affiliation{Quantum Electromagnetics Division, National Institute of Standards and Technology, Boulder, CO 80305, USA}
 
\author{Hans T. Nembach} 
\affiliation{Department of Physics and JILA, University of Colorado and NIST, Boulder, CO 80309, USA}
 \affiliation{Quantum Electromagnetics Division, National Institute of Standards and Technology, Boulder, CO 80305, USA}
 
\author{Yaroslav Kvashnin} 
 \affiliation{Department of Physics and Astronomy, Uppsala University, Box 516, 75120 Uppsala, Sweden}
  
  \author{Robert Stefaniuk} 
 \affiliation{Department of Physics and Astronomy, Uppsala University, Box 516, 75120 Uppsala, Sweden}
 
  \author{Rameez S. Malik} 
 \affiliation{Department of Physics and Astronomy, Uppsala University, Box 516, 75120 Uppsala, Sweden}

\author{Patrik Grychtol} 
\affiliation{Department of Physics and JILA, University of Colorado and NIST, Boulder, CO 80309, USA}

\author{Dmitriy Zusin} 
\affiliation{Department of Physics and JILA, University of Colorado and NIST, Boulder, CO 80309, USA}

\author{Christian Gentry} 
\affiliation{Department of Physics and JILA, University of Colorado and NIST, Boulder, CO 80309, USA}

\author{Raghuveer Chimata} 
 \affiliation{Department of Physics and Astronomy, Uppsala University, Box 516, 75120 Uppsala, Sweden}
 
 \author{Manuel Pereiro} 
 \affiliation{Department of Physics and Astronomy, Uppsala University, Box 516, 75120 Uppsala, Sweden}

\author{Johan Soderstrom} 
 \affiliation{Department of Physics and Astronomy, Uppsala University, Box 516, 75120 Uppsala, Sweden}

\author{Emrah Turgut}
\affiliation{Department of Physics and JILA, University of Colorado and NIST, Boulder, CO 80309, USA}

\author{Martina Ahlberg} 
\affiliation{Department of Physics, University of Gothenburg, 412 96 Gothenburg, Sweden}

\author{Johan {\AA}kerman} 
\affiliation{Department of Physics, University of Gothenburg, 412 96 Gothenburg, Sweden}

\author{Henry C. Kapteyn} 
\affiliation{Department of Physics and JILA, University of Colorado and NIST, Boulder, CO 80309, USA}

\author{Margaret M. Murnane} 
\affiliation{Department of Physics and JILA, University of Colorado and NIST, Boulder, CO 80309, USA}

\author{D. A. Arena} 
\affiliation{Department of Physics, University of South Florida, Tampa, FL 33620, USA}

\author{Olle Eriksson} 
 \affiliation{Department of Physics and Astronomy, Uppsala University, Box 516, 75120 Uppsala, Sweden}

\author{Olof Karis} 
 \affiliation{Department of Physics and Astronomy, Uppsala University, Box 516, 75120 Uppsala, Sweden}

\author{T. J. Silva} 
 \affiliation{Quantum Electromagnetics Division, National Institute of Standards and Technology, Boulder, CO 80305, USA}

\date{\today}

\begin{abstract}
Ni$_{0.8}$Fe$_{0.2}$ (Py) and Py alloyed with Cu exhibit intriguing ultrafast demagnetization behavior, where the Ni magnetic moment shows a delayed response relative to the Fe, an effect which is strongly enhanced by Cu alloying. 
We have studied a broad range of Cu concentrations to elucidate the effects of Cu alloying in Py. 
The orbital/spin magnetic moment ratios are largely unaffected by Cu alloying, signifying that Cu-induced changes in the ultrafast demagnetization are not related to spin-orbit interactions. We show that magnon diffusion can explain the delayed Ni response, which we attribute to an enhanced magnon generation rate in the Fe sublattice relative to the Ni sublattice. Furthermore, Py exhibits prominent RKKY-like exchange interactions, which are strongly enhanced between Fe atoms and diminished between Ni atoms by Cu alloying. An increased Fe magnon scattering rate is expected to occur concurrently with this increased Fe-Fe exchange interaction, supporting the results obtained from the magnon diffusion model.
 \end{abstract}

\maketitle

\section{Introduction}

The results of ultrafast demagnetization remains challenging to fully understand, especially from a theoretical standpoint, where non-equilibrium conditions and complex interactions between electronic spins, phonons and magnons need to be considered. 
Even for simple metal multilayer stacks, the contributions from Elliot-Yafet spin flips\cite{Carva_PRB_2013, Koopmans_NatMat_2010} and superdiffusive spin currents\cite{PhysRevB.86.024404} to the demagnetization are difficult to determine. Furthermore, mechanisms such as magnon generation\cite{PhysRevB.90.014417} and feedback effects\cite{PhysRevLett.111.167204} may also play an important role.
 During demagnetization, a large number of electrons near the Fermi level are excited, and it has been shown that low energy electrons in Fe scatter strongly with magnons\cite{PhysRevB.62.5589}. This is in accordance with experimental findings\cite{PhysRevB.78.174422, PhysRevLett.105.197401,PhysRevB.91.174414, PhysRevLett.121.087206} and theoretical studies that show ultrafast magnon generation dominating over phonon mediated spin-flip scattering\cite{PhysRevB.90.014417} in Fe. Ultrafast magnon generation is also expected in Ni\cite{PhysRevB.90.014417}. However, the demagnetization of Ni is also accompanied by a decrease in exchange spitting\cite{Rhie2003247201/1,Tengdineaap9744}, which is a signature of Stoner-like excitations.
 Increasing evidence suggests that Co, like Fe, demagnetizes primarily through ultrafast magnon generation\cite{PhysRevB.94.220408,Eiche1602094,Cinchetti_PRL_2006}. During demagnetization, excited electrons generate magnons through exchange scattering, meaning that an uncorrelated minority electron scatters into the majority band when a (correlated) magnon is generated, leaving the net magnetization unchanged\cite{PhysRevB.90.014417}. Hence, magnon generation is necessarily a secondary effect, concurrent with a demagnetizing process such as Elliot-Yafet spin-flip scattering or superdiffusive spin currents. This undetermined demagnetization process will here be denoted itinerant spin-flip, to distinguish it from magnons that are associated with localized magnetic moments.
For elemental materials (non-alloyed) the subsequent magnon dynamics are irrelevant during demagnetization, since the magnon has equal probability to be found at any lattice site at all times. However, for alloys, 
it is only when the alloying elements are indistinguishable, such as in the rigid band model, that  an identical magnon probability for all lattice sites can be ensured,  which is not the case for transition metal alloys\cite{Zeller19872123}.
Ultrafast magnetization dynamics between elements in an alloy has been modeled by Schellekens et al.\cite{PhysRevB.87.020407} and Hinzke et al.\cite{PhysRevB.92.054412}, but neither study takes into account the fundamental magnon excitation and its time evolution.
Due to the significantly different propensity for magnon generation between Fe and Ni\cite{PhysRevB.90.014417}, it can be expected that Fe-Ni alloys should be ideal for studying the effect of inhomogeneous magnon scattering during demagnetization. We will show that a direct consequence of this phenomena was observed by Mathias et al.\cite{Mathias20124792} when demagnetizing Permalloy Ni$_{0.8}$Fe$_{0.2}$ (Py), which has been reproduced by others\cite{Somnath_RevSciInst_2017, PhysRevB.90.180407}. It was found that Ni exhibits an 18 fs delay in its demagnetization relative to Fe. Furthermore, they found that by alloying with 40\% Cu the delay increased to 76 fs. They related the delayed response of Ni to the average exchange interaction of the alloy. However, the microscopic mechanism was undetermined.

To understand the significantly increased delay of Ni in Py$_{0.6}$Cu$_{0.4}$, it is necessary to study how Cu alloying affects the magnetic properties of Py.  
We present a study on a series of (Ni$_{0.8}$Fe$_{0.2}$)$_{1-x}$Cu$_x$ (i.e. Py$_{1-x}$Cu$_{x}$) alloys, to determine the effect of Cu alloying on the magnetization and exchange interactions. Using X-ray magnetic circular dichroism (XMCD), we show that the orbital/spin ratio is only marginally affected by Cu alloying, eliminating 
the possibility for spin-orbit related effects as the cause for the increased delay in Py$_{0.6}$Cu$_{0.4}$. The Ni spin magnetic moment is found, using XMCD and ab-inito calculations, to decrease significantly with Cu alloying while Fe remains largely unaffected. This is concurrent with a significantly decreasing RKKY-like exchange interaction between Ni atoms, however, the RKKY-like exchange interaction is found to strongly increase between Fe atoms.
Using a magnon diffusion model (MDM) we show that the remagnetization time of Fe, Ni and Py correspond well to what can be expected from Gilbert damping of magnons. It is found that the delayed demagnetization of Ni relative to Fe in Py$_{1-x}$Cu$_{x}$ is well described by an enhanced magnon generation rate at Fe sites relative to Ni sites. Furthermore, the model indicates that this asymmetry in magnon generation rate is increased by Cu alloying. This is well correlated to the increased RKKY-like exchange interaction that should result in an increased magnon scattering rate in the Fe sublattice.

%This is due to changes in the RKKY-like exchange interaction, which increases between Fe atoms and decreases between Ni atoms with increasing Cu concentration.

\section{Elemental magnetization}
\begin{figure}
	\begin{center}
          \includegraphics[width=0.47\textwidth]{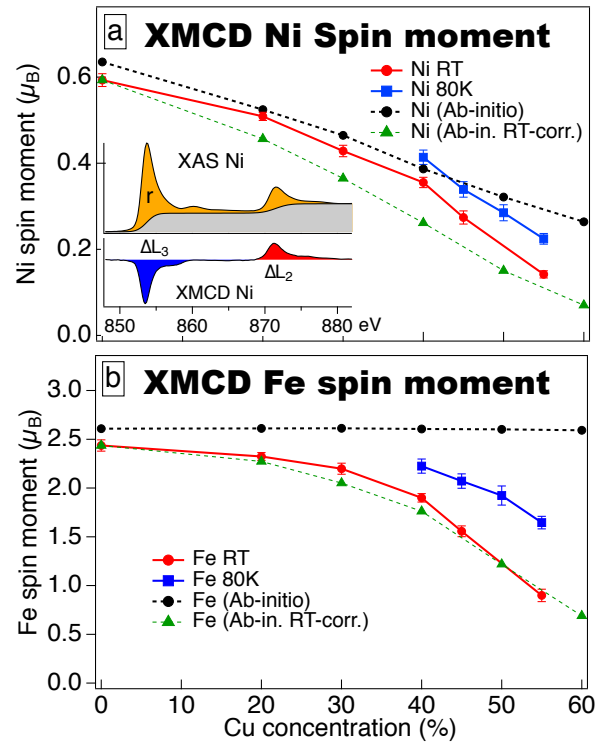}
           \end{center}
      \caption{\label{Spin_FeNi_ratio} (Color online) Spin magnetic moment of Ni (a) and Fe (b) as a function of Cu concentration. The theoretical (black dashed) Ni moment in (a) decreases continuously with Cu concentration while the Fe moment in (b) remains constant. Inset in (a) shows the absorption and XMCD of Ni in Py to illustrate the parameters used in Eq.\ \ref{eq1}.} 
\end{figure}

Element-selective spin and orbital magnetic moments are here studied by x-ray magnetic circular dichroism (XMCD) and ab-initio calculations to determine how Cu alloying affects the magnetic and electronic properties in a series of Py$_{1-x}$Cu$_{x}$ alloys.
The spin and orbital magnetic moments are given by the XMCD sum rules\cite{PhysRevLett.75.152} and are proportional to the number of 3d-holes ($N_h$), 
\begin{equation}
\begin{aligned}
m_s=& -\frac{\Delta L_3 - 2 \Delta L_2}{r}\cdot  N_h \\
m_l=& -\frac{2}{3} \frac{\Delta L_3 + \Delta L_2}{r}\cdot  N_h
\label{eq1}
\end{aligned}
\end{equation}
The parameter $r$ is the total integrated intensity of the absorption edge (averaged over both magnetization directions) after subtracting a background (gray are in Fig\ \ref{Spin_FeNi_ratio}a, inset). The position of the steps in the background were taken at the energies corresponding to the peak absorption of the L$_3$ and L$_2$ edges with a step height ratio of 2:1. $\Delta L_n$ ($n=2,3$) are the integrated absorption differences between M+ and M- spectra at the $L_n$ absorption edges, as illustrated in Fig.\ \ref{Spin_FeNi_ratio}a inset. The magnetic dipole term can safely be ignored since the sample is thick enough to make any interface effects negligible\cite{PhysRevApplied.2.044014}. The spin moments of Ni and Fe obtained from XMCD and ab-initio calculations are presented in Fig.\ \ref{Spin_FeNi_ratio}a and b, respectively, for different Cu concentrations. The 0 K ab-initio spin moments are shown as black dashed/circles. The dashed line/triangles correspond to ab-initio values after correcting for a reduced magnetization at room temperature. The room temperature (RT) correction was obtained from SQUID magnetometry, where we measured the ratio in total magnetization between 300 K and 10 K. This temperature correction is strictly valid only if Fe and Ni have equivalent temperature dependence, which has been shown to be the case for Py$_{0.6}$Cu$_{0.4}$ (see SI in Ref.\ \cite{Mathias20124792}).
To extract the XMCD spin moments, we used the number of holes in Eq.\ \ref {eq1} as a fitting parameter to equate the spin magnetic moment from XMCD and the ab-initio calculations at RT for Permalloy (Py, 0\% Cu), giving 3.41 holes for Fe and 1.28 holes for Ni in Py, which are slightly lower than the ab-initio values of 3.47 holes for Fe and 1.40 holes for Ni.

Our ab-initio spin moments for Py are consistent with values reported by neutron scattering \cite{Collins_someTAlloys}, where the magnetic moment for Ni is similar to the pure element ($\sim 0.6 \mu_B$) but the magnetic moment of Fe is enhanced by approximately 20\% relative to that of pure bcc Fe.
The theoretical 0 K Ni spin moment in Fig.\ \ref{Spin_FeNi_ratio}a exhibit a roughly linear decrease with increasing Cu concentration, but remains non-zero even up to 90\% Cu. 
The RT XMCD Ni moment shows values that are somewhat higher than what is obtained from RT-corrected ab-initio values.
Surprisingly, for the Fe spin moment shown in Fig.\ \ref{Spin_FeNi_ratio}b, theory suggests an almost constant value that persists all the way to 100\% Cu at $T=0$ K. 
The RT XMCD measurements follow closely the RT-corrected ab-initio values, supporting the 0 K ab-initio result that the local Fe magnetic moment is virtually unaffected by Cu alloying.

 \begin{figure}
	\begin{center}
          \includegraphics[width=0.53\textwidth]{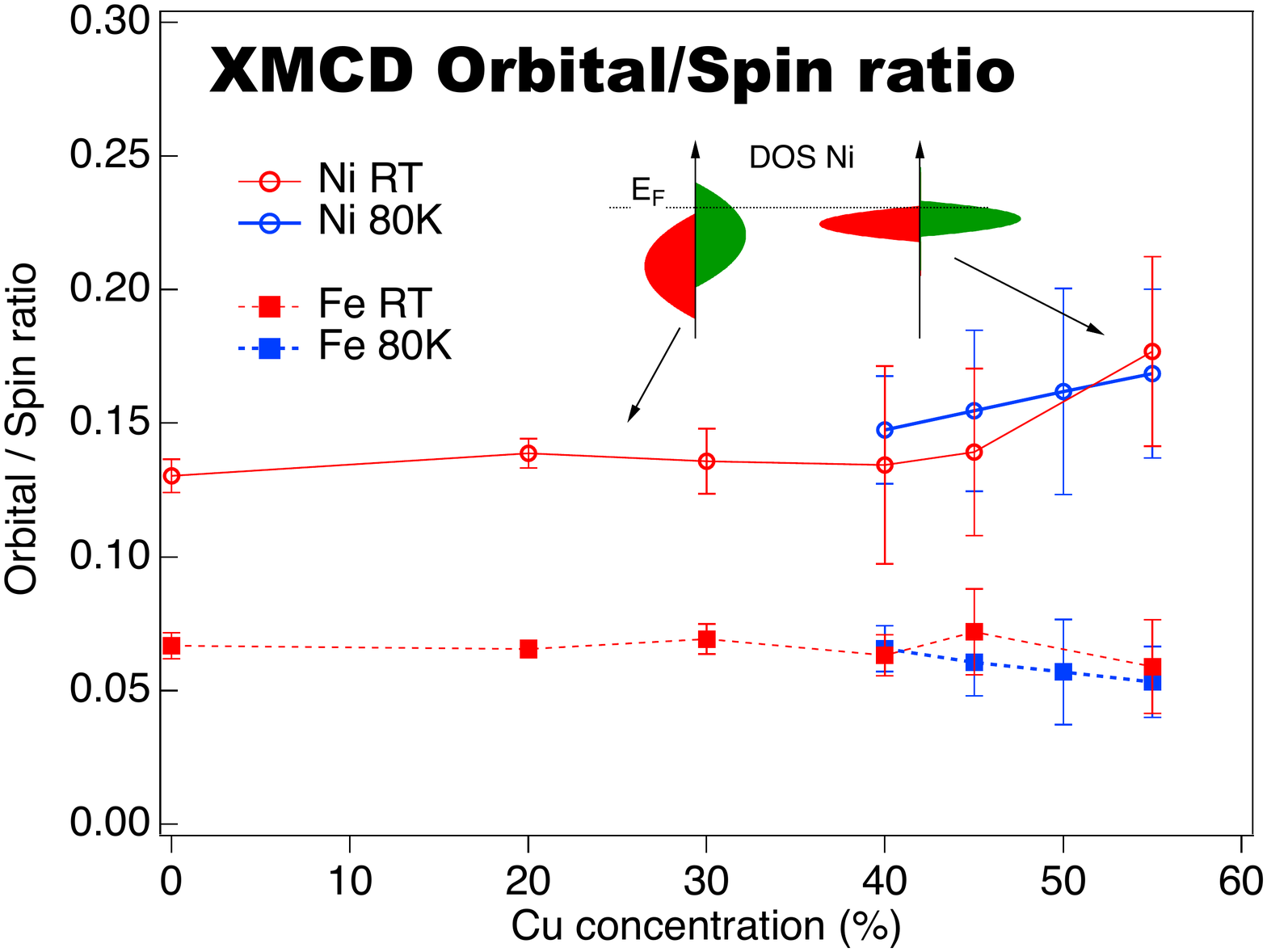}
           \end{center}
      \caption{\label{Spin_Orbit} (Color online) The ratios between the orbital and spin moments for Fe and Ni in Py$_{1-x}$Cu$_x$ as a function of Cu concentration. The ratio remains constant for low concentrations of Cu alloying. At high Cu concentrations ($>40$\%) the ratio increases for Ni.}
\end{figure}

The orbital moment has a very similar dependence on Cu concentration as the spin magnetic moment.
To identify any Cu concentration dependent trends in the orbital moment, we studied the orbital/spin ratio, shown in Fig.\ \ref{Spin_Orbit}. Unlike the spin and orbital moment by themselves, the $m_l/m_s$ ratio is independent of the number of $3d$-holes, as apparent from Eq.\ \ref{eq1}. The values are slightly higher than what has been measured in other studies for bcc Fe (0.043)\cite{PhysRevLett.75.152}, fcc Ni (0.11)\cite{PhysRevB.53.3409} and fcc Py (Ni: 0.10, Fe: 0.04)\cite{Glaubitz2011}. The ratio for both Fe and Ni remains almost constant through a wide range of Cu concentrations. However, there appears to be an increase in the ratio for Ni at high Cu concentrations which can be related to the narrowing of occupied Ni $d$-states as a consequence of the decreased Ni $d$-$d$ overlap. See inset in Fig.\ \ref{Spin_Orbit}. 

\section{Exchange interactions}
\begin{figure}
	\begin{center}
          \includegraphics[width=0.52\textwidth]{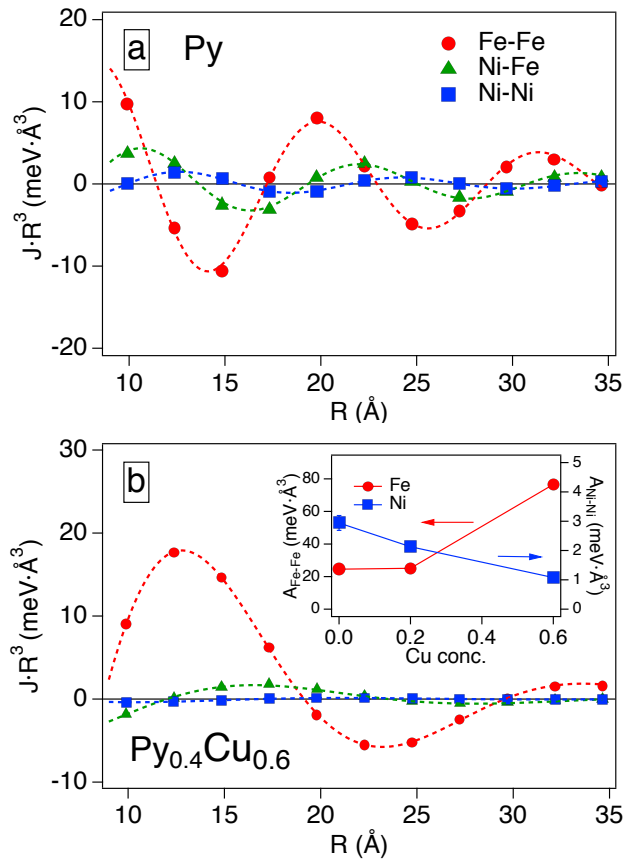}
           \end{center}
      \caption{\label{RKKY_PyCu} (Color online) a) Exchange interaction scaled with R$^3$ between Fe-Fe, Fe-Ni and Ni-Ni atoms in Py. Dashed lines are fits to a damped harmonic oscillation. b) Exchange interactions for Py$_{0.4}$Cu$_{0.6}$. Inset: Fitted amplitudes of the oscillations for different Cu concentrations for Ni-Ni (blue squares) and Fe-Fe (red circles).}
\end{figure}

The magnon generation rate is proportional to the square of the exchange interaction between free and localized magnetic moments $(J_{sd})^2$  (see Methods). An easily accessible method to determine changes in $J_{sd}$ is to investigate the amplitude of RKKY-like exchange interactions which are also proportional to $(J_{sd})^2$.
It has been found that the exchange interaction in 3$d$-elements exhibits an RKKY-like interaction\cite{PhysRevB.64.174402}, which for Fe has been related specifically to t$_{2g}$ levels\cite{PhysRevLett.116.217202}. RKKY-like exchange interactions are found in Py$_{1-x}$Cu$_x$ alloys, as presented in Fig.\ \ref{RKKY_PyCu}, which shows $JR^3$ along the (110) direction, where $J$ is the exchange interaction and $R$ is the inter-atomic distance. The data has been fitted with $JR^3=A\cdot sin(k\cdot R+\theta) e^{(-R/l)}$, where $A$ is the amplitude of the oscillation, $k$ is the wave number, $\theta$ is the phase and $l$ is the decay length of the interaction\cite{PhysRevB.64.174402}. Fig.\ \ref{RKKY_PyCu}a shows the exchange interaction between Fe-Fe, Ni-Ni and Fe-Ni in Py. The interaction is damped with $l\approx1.8$ nm for all elements in Py. The amplitude is about an order of magnitude higher between Fe atoms than between Ni atoms in Py, as can be seen in the inset of Fig.\ \ref{RKKY_PyCu}b. Alloying with $60\%$ Cu decreases the decay length to $l\approx1.0$ nm, while the amplitude of the RKKY interaction between Fe atoms increases by a factor 3 and the Ni-Ni interaction decreases, which is consistent with a decreasing Ni magnetic moment. The wave number $k$ obtained from the fits are 5.4, 4.9, and 3.0 nm$^{-1}$ between all elements in Py, Py$_{0.8}$Cu$_{0.2}$ and Py$_{0.4}$Cu$_{0.6}$, respectively.

Experimentally, it has been shown that Fe-Cu alloys can indeed exhibit spin-glass behavior\cite{PhysRevB.33.3247}, which is indicative of such RKKY-like exchange interactions\cite{Mohn}. 

\section{Remagnetization}

 \begin{figure}
	\begin{center}
          \includegraphics[width=0.5\textwidth]{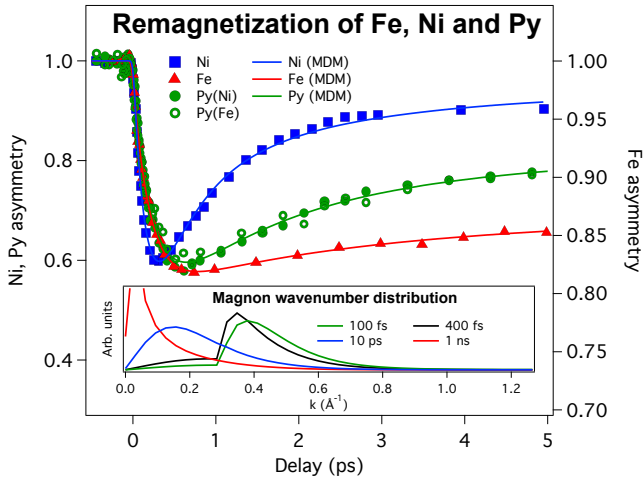}
           \end{center}
      \caption{\label{Demag_Fe_Ni_Py} (Color online) Ultrafast remagnetization of bcc Fe (triangles), fcc Ni (squares) and fcc Py (circles). Solid lines are fits using the magnon diffusion model. For Py, both Fe (hollow circles) and Ni (solid circles) asymmetries are presented, while the fit corresponds to the average magnetization. The inset shows the magnon $k$-distribution after various time delays in Py.}
\end{figure}

If the electron system has thermalized and cooled down, the remagnetization should depend on the magnon damping of the system. 
Several theoretical studies on ultrafast demagnetization attempt to relate the demagnetization to the Gilbert damping parameter\cite{0295-5075-81-2-27004,PhysRevLett.95.267207,PhysRevB.92.054412,PhysRevB.84.144414}. Here we instead focus on the remagnetization of three samples, Fe, Ni and Py as shown in Fig.\ \ref{Demag_Fe_Ni_Py}. 
The demagnetization times are relatively similar for all three samples (see Table I), while the remagnetization varies strongly.
The solid lines corresponds to fits obtained from MDM and where the fitting parameters are the onset of demagnetization, demagnetization amount, demagnetization time, remagnetization amount and the Gilbert damping. The MDM is described in the methods section. The demagnetization is modelled by a single exponential that generates magnons with a distribution of wavenumbers ($k_0$) presented in the inset of Fig.\ \ref{Demag_Fe_Ni_Py}, for various time delays in Py. At long timescales (1 ns) the magnon distribution resembles a Bose-Einstein distribution. However, at short time scales, the high-k magnons with shorter lifetimes dominate. Here we use the distributions obtained at 400 fs, which is approximately when the remagnetization becomes apparent. 
The electron temperature is unknown and expected to be strongly varying during the first picosecond, however, we have used a constant value of 700 K for all cases which can be considered a reasonable average value during the demagnetization.\cite{Rhie2003247201/1}
The magnons that are generated during demagnetization are subject to Gilbert damping with a time constant of $\tau_{damp}=\frac{1}{2\alpha_G\cdot \omega (k)}$, where $\alpha_G$ is the Gilbert damping parameter and $\omega$ is magnon angular frequency.
The fitted values of $\alpha_G$ for Fe and Py are $2.4\cdot10^{-3}$ and $3.9\cdot10^{-3}$, respectively. These values compare well with values obtained from FMR\cite{M_Schoen_PRB_2017} of $2.5\cdot 10^{-3}$ for Fe and $5.0\cdot 10^{-3}$ for Py.
See Table \ref{FittingParam} for all fitting parameters. The Ni does not converge within reasonable value of the fitting parameters, likely due to the similar timescales of demagnetization and remagnetization, therefore the Gilbert damping parameter was fixed at $2.4\cdot10^{-2}$ as given in Ref. \cite{M_Schoen_PRB_2017}. 

 The strong correlation between MDM and FMR values supports the notion that magnons are generated on ultrafast timescales, but remagnetization is moderated by conventional Gilbert damping. This naturally raises the question why magnon emission during demagnetization is so much faster than the magnon absorption during remagnetization. Fast magnon generation results from the interband transition of itinerant spins, where a minority electron scatters into the majority band, when a magnon is created, effectively conserving the total magnetization, i.e.,  Landau scattering\cite{PhysRevLett.109.087203}. Therefore the demagnetization is still ultimately limited by the itinerant spin-flip rate. However, the net flow of angular momentum to the lattice can be enhanced by the continuous transfer of minority electrons to the majority band by fast magnon generation\cite{PhysRevB.90.014417}. It was shown by Carva et al.\cite{Carva_PRB_2013} that the EY-spin-flip rate can be orders of magnitude faster for an optically generated far-from-equilibrium electron distribution, as compared to the case of thermalized low temperature distribution.  The electron temperature is expected to approach the lattice temperature within approximately 1 ps\cite{Rhie2003247201/1, PhysRevB.78.174422}. It is therefore possible for the magnon decay rate induced by the Gilbert damping to completely dominate over the EY-spin-flip rate during the remagnetization process. 

\section{Simulation of inhomogeneous magnon scattering in alloys}
In the above section there was no distinction between different lattice sites. Here we generate a 1D lattice of a random alloy where the Fe and Ni sites have different propensity for magnon generation. 
The main objective is to determine if ultrafast and efficient magnon generation in the Fe sublattice can explain the delayed demagnetization of the Ni sublattice. 
Details of the MDM are described in the Methods section. A 1D lattice with 1024 atoms is generated and every atom is randomly assigned as Fe and Ni with a probability corresponding to the concentration of each element. 
The parameter $p$ is used for describing the ratio in the probability of finding a magnon at an Fe site relative to a Ni site due to the difference in exchange scattering ($p \sim (\frac{J^{sd}_{Fe}}{J^{sd}_{Ni}})^2$, see Methods).

\begin{figure}
	\begin{center}
          \includegraphics[width=0.50\textwidth]{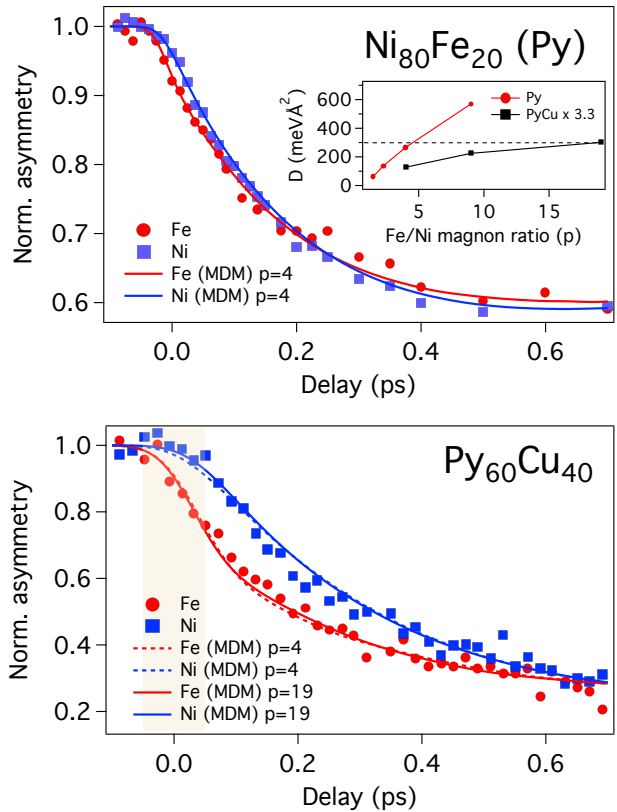}
           \end{center}
      \caption{\label{Demag_Py_Cu} (Color online) a) Demagnetization of Py. Fe (red, circle) demagnetizes before Ni (blue, square). Fit with magnon diffusion model (solid lines). Inset: Fitted values of the spin wave stiffness for various values of the parameter $p$. b) Demagnetization of Py$_{0.6}$Cu$_{0.4}$ adopted from Mathias et al.\cite{Mathias20124792}. Fits are with two different values of Fe scattering probability, $p$=19 (solid lines) and $p$=4 (dashed lines).}
\end{figure}
Fig.\ \ref{Demag_Py_Cu}a shows the elementally resolved asymmetry of Py during demagnetization. As found by Mathias et al.\cite{Mathias20124792}, the Fe (red circles) appears to have a faster demagnetization response at $t=0$ than Ni. The experimental data has been fitted using the diffusion model with $p=4$ (solid lines) and a preferential weighting between $-40<t<100$ fs, to ensure that the delayed response of Ni relative to Fe is properly captured. The extracted values of the spin-wave stiffness ($D$) for various values of the parameter $p$ is plotted in the inset. The general trend is that $p$ increases if $D$ increases. This is due to increased dispersion when $D$ increases which decreases the delay between Fe and Ni, this effect is compensated by an increased $p$. The horizontal dashed line corresponds to the experimentally determined value of the spin-wave stiffness ($\sim$300 meV\AA$^2$) obtained from neutron scattering \cite{Vac_RepProgPhys_2008}. 
The fit that is nearest to $D=300$ meV\AA$^2$ is obtained for $p=4$ which provides a spin-wave stiffness ($D$) of $267\pm21$ meV\AA$^2$. This would indicate that the s-d exchange interaction is a factor $\sim$2 higher for Fe compared to Ni in Py.
In Fig.\ \ref{Demag_Py_Cu}b we show experimental data adopted from Ref.\ \cite{Mathias20124792} for demagnetization of Py$_{0.6}$Cu$_{0.4}$. Unlike Py, the Py$_{0.6}$Cu$_{0.4}$ shows a wide time interval $\sim 100$ fs, where Fe demagnetizes but Ni does not (see yellow shaded region). Fits using $p=4$ (dashed lines) gives a value of $D=39\pm2$ meV\AA$^2$ for the exchange stiffness. 
The exchange energy, and hence the also the spin wave stiffness, is expected to be 3.3 times less\cite{Mathias20124792} in Py$_{0.6}$Cu$_{0.4}$ compared to Py ($\sim 300/3.3\approx 91$ meV\AA$^2$) which is higher than the $D$ value obtained when $p=4$. The solid lines corresponds to fits with $p=19$ that gives $D=91\pm7$ meV\AA$^2$, which is the expected value. This would indicate that the s-d exchange interaction is a factor $\sim$4 higher in Fe compared to Ni in Py$_{0.6}$Cu$_{0.4}$, which is a factor $\sim$2 higher ratio than for Py and is consistent with the increase in RKKY-like exchange interactions with Cu alloying.
The damping is expected to be about 3 times higher in Py$_{0.6}$Cu$_{0.4}$ than Py\cite{Mathias20124792}, hence we have used a fixed value of $\alpha_G=0.01$. See Table \ref{FittingParam} for all fitting parameters. 

\section{Discussion}
 
The large delay of Ni demagnetization in Py$_{0.6}$Cu$_{0.4}$, as shown in Fig.\ \ref{Demag_Py_Cu}b, could in principle be explained by a vanishing spin-flip probability in the Ni sublattice. Schellekens and Koopmans\cite{PhysRevB.87.020407} showed that a factor 4 smaller spin-flip probability in Ni compared to Fe could explain the demagnetization in Py. However, using that model to describe Py$_{0.6}$Cu$_{0.4}$ demagnetization would give unreasonably small values of the Ni spin-flip probability approaching zero. It follows, if the spin flip probability is kept within reasonable values, that fast angular momentum transfer between the Ni and Fe sublattices is required in order to explain the experimental data. Theoretically it has been shown by Haag, Illg and F{\"a}hnle\cite{PhysRevB.90.014417, Haag2013, PhysRevB.88.214404} that ultrafast magnon generation is very efficient in elemental Fe, while it is an order of magnitude slower in Ni. Furthermore, this process is potentially more efficient than the electron-phonon spin-flip scattering process. It was also suggested that  magnon scattering can result in enhanced electron-phonon scattering rates, bridging the gap between the low theoretically calculated EY spin flip probability and experimental demagnetization rates. 
Experimentally there are several indirect measurements indicating the existence of ultrafast magnon generation during demagnetization\cite{PhysRevB.94.220408,PhysRevLett.105.197401,PhysRevB.78.174422}. 
This suggests that a qualitatively accurate description of ultrafast demagnetization should include the effect of electron-magnon scattering.

\begin{figure}
	\begin{center}
          \includegraphics[width=0.52\textwidth]{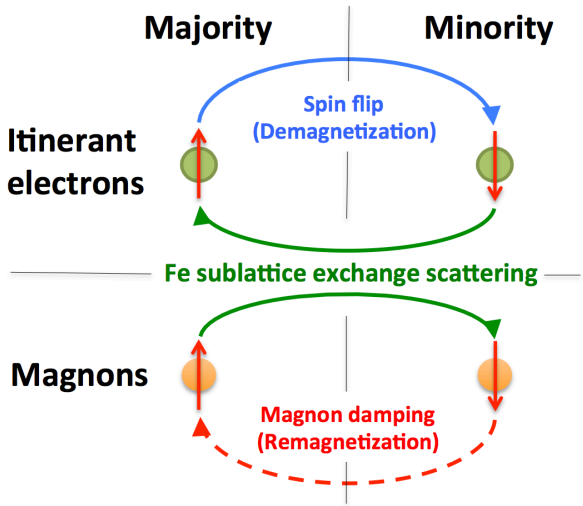}
           \end{center}
      \caption{\label{SpinFlipMagnon} (Color online) Illustration of magnon generation and spin flip processes during demagnetization. Itinerant spin-flips transfer angular momentum from the spin system which drives the demagnetization. An efficient magnon scattering (emission) in the Fe sublattice is effectively transferring angular momentum between delocalized magnetic moments to localized Fe moments, enhancing the demagnetization of Fe and subsequently diminishing the demagnetization of Ni. Remagnetization is determined by the Gilbert damping.}
\end{figure}
We propose that a large difference in the magnon scattering efficiency between the Fe and Ni sublattices could explain the anomalous demagnetization in both Py and Py$_{0.6}$Cu$_{0.4}$. The demagnetization process is illustrated in Fig.\ \ref{SpinFlipMagnon}, where the angular momentum is transferd from the spin system through an itinerant spin-flip process that drives the demagnetization. At the same time, there is efficient magnon generation in the Fe sublattice. Magnon generation does not change the total magnetization since a minority spin is flipped concurrently, so effectively the magnon generation will increase the demagnetization of the Fe sublattice on the expense of a decreased demagnetization in the Ni sublattice. The magnons will decay on much longer time scales, corresponding to the Gilbert damping, which defines the behavior of the remagnetization. We have shown that the remagnetization of Fe, Ni and Py is well described by their respective Gilbert damping parameters.  In section VI, we model the diffusion of magnons that are preferentially generated in the Fe sublattice. We find that the fitted values of the spin-wave stiffness agree well with known experimental values. For Py$_{0.6}$Cu$_{0.4}$ we find that a substantial increase of the magnon generation efficiency in the Fe sublattice, compared to Py, is necessary for fitting to reasonable values of the spin-wave stiffness.
This correlates to a strongly enhanced RKKY-like exchange interaction in the Fe sublattice when Py is alloyed with Cu. 
Future ab-initio calculations, where the magnon generation has been implemented in a framework for alloys, should be able to substantiate the assumption of preferential magnon scattering in the Fe sublattice of Py. 

We note that it is a common misconception that results by Eschenlohr\cite{EschenlohrThesis}  and Radu et al.\ \cite{doi:10.1142/S2010324715500046} contradict the findings of a delayed Ni response in Py. However, as explicitly explained in Ref.\ \cite{EschenlohrThesis}, a large inherent uncertainty in the determination of the pump-pulse arrival time was handled by shifting the Fe and Ni data sets to eliminate any time-delay between Fe and Ni, and hence the above-mentioned effect could not be addressed by those measurements.

The magnon diffusion model has been corroborated by a systematic study of demagnetization in a concentration series of Fe$_{x}$Ni$_{1-x}$ alloys by S. Jana et al.\cite{Somnath_FeNi_alloy}. 

\section{Acknowledgments}

We thank Yves Idzerda, Karel Carva and Peter Oppeneer for useful discussions. The authors are grateful to Hermann D{\"u}rr for the initial suggestion to study PyCu orbital moments with XMCD.
This work was partially supported by the U.S Department of Energy Office of Basic Energy Sciences X-Ray Scattering Program (Grants \#DE-SC0002002 and \#DE-FG02-09ER46652).
O.E. acknowledges support from the KAW foundation and VR. R.K. acknowledges the Swedish Research Council (VR) for their financial support. E. K. D.-Cz. acknowledges National Supercomputer Centre at Link{\"o}ping University Sweden for computational resources and STandUP for financial support. P.G. acknowledges support from the Deutsche Forschungsgemeinschaft (no. GR 4234/1-1). Use of the National Synchrotron Light Source, Brookhaven National Laboratory, was supported by the U.S. Department of Energy, Office of Science, Office of Basic Energy Sciences, under Contract No. DE-AC02-98CH10886.

\section{Methods}

\subsection {X-ray magnetic circular dichroism}
The XMCD measurements were performed at beamline U4B at the National Synchrotron Light Source at Brookhaven National Laboratory in (Upton, NY), using 70\% circularly polarized light at normal incidence to the samples. The samples were saturated out-of-plane (OOP) using a magnetic field of 1.5 T. The magnetic contrast was obtained by switching the OOP magnetic field, where the corresponding spectra are referred to as M$+$ and M$-$. The measurements were performed in transmission. 
Unlike total electron yield (TEY) measurements, which are more commonly used, the transmission geometry is insensitive to externally applied fields and is not affected by saturation effects which can affect the derived magnetic moments\cite{PhysRevB.59.6421}. Also, transmission measurements provide a much more reliable method for comparing the absorption coefficient between different samples, as they are insensitive to sample dependent electron escape depth and generation of secondary electrons\cite{PhysRevB.59.6421}. A reference sample without the Cu-Py layer was used to correct for the absorption from the Si$_3$N$_4$ and Ta layers.  The provided uncertainties correspond to the standard deviation and are derived by estimating the variance in the difference and sum signal of the M$+$ and M$-$ spectra. Also, an estimated uncertainty in the fitting of the step function was included in the total variance. The accuracy of the film thickness and Cu concentration will affect the analysis of the total absorption cross-section, and was estimated to have a standard deviation of 3\%, as determined from x-ray reflectivity measurements.
For the XMCD study, thin films of 20 nm (Fe$_{0.2}$Ni$_{0.8}$)$_{1-x}$Cu$_x$, with $0<x<0.55$ were grown on 3 nm Ta seed layers by magneton sputtering on Si$_3$N$_4$ substrates. The samples were capped by 3 nm Ta to prevent oxidation. Two sample sets, S1 and S2, were grown at two different occasions in the same sputtering system. The results from both sample sets were practically identical, confirming the repeatability of our procedure. Here we present the averaged data of the two samples sets. Our samples were grown in the same sputtering system and with the same growth parameters as for the samples studied by Mathias et al.\cite{Mathias20124792} (see their supporting information for sample characterization). 

\subsection {Ultrafast demagnetization}
Ultrafast demagnetization of Py was measured at the high harmonic generation source HELIOS using the transverse magneto-optical Kerr effect (T-MOKE); see Ref.\cite{Somnath_RevSciInst_2017} for specifications of this source. The sample geometry used for ultrafast demagnetization was Si-substrate/Ta(2 nm)/X(20 nm)/Ta(2 nm), where X is Py, Fe or Ni. The pump fluence for Py, Ni and Fe were 1.4 mJ/cm$^2$, 1.8 mJ/ cm$^2$, and Fe 7.2 mJ/cm$^2$ respectively. The presented Py$_{0.6}$Cu$_{0.4}$ data are adopted from Ref.\cite{Mathias20124792} where a similar source was used.
 Table I shows the parameters and values obtained from fitting the experimental data. The error margin is the standard deviation for 5 simulations with different random positions for Fe and Ni atoms. 

\begin{table}
 \caption{Parameters from the fittings in Figs.\ \ref{Demag_Fe_Ni_Py} and \ref{SpinFlipMagnon}. The superscript $^f$ indicates a fixed parameter. $\tau_{dis}$ is the time at which the magnon $k_0$ distribution is obtained (ps), $p$ is the probability of magnon generation in the Fe lattice, D is the spin-wave stiffness (meV\AA$^2$), $\alpha_{G}$ is the Gilbert damping parameter,  $\tau_{dem}$ is the demagnetization time (ps), $A_{end}$ is the fraction of magnons that become damped, and $M_{end}$ is the amount of demagnetization. }
\begin{tabular}{cccccccc}
\firsthline
\label{FittingParam}

Sample    		        & $\tau_{dis}$   & $p$ 		& $D$   & $\alpha_{G} (10^{-3})$ 	& $\tau_{dem}$ & $A_{end}$ & $M_{end}$ \\
\hline
Fe      			& 0.4$^f$		&   -   		& 270$^f$   & 2.4  				& 0.200  	&		0.35		&  0.20 \\
Ni        			& 0.4$^f$		&   -   		& 400$^f$   & 24$^f$ 			& 0.176 	&		0.95  	&  0.69	 \\
Py       			& 0.4$^f$		&   1$^f$  	& 320$^f$   & 3.9  				& 0.237	&		0.67 		&  0.51 \\
           			& 0.1$^f$		&   4$^f$    	& 267   	  & 3.1 				& 0.233	&		0.66 		&  0.50  \\
Py$_{60}$Cu$_{40}$	 &  0.1$^f$ 	&  4$^f$		& 39		  & 10$^f$				& 0.328	&		0.33$^f$	& 0.84 \\
	   			&  0.1$^f$ 	&  19$^f$ 	& 91	  	  & 10$^f$				& 0.333	&		0.33$^f$	 & 0.88  \\
Std dev. 			& -			& 	-		& $\pm$8\%	  & $\pm$6\%		& $\pm$2\%		& $\pm$2\%  & $\pm$2\% \\
\lasthline
\end{tabular}
\end{table}

\subsection {Ab-initio calculations}
Ab-initio calculations were performed using the spin-polarized relativistic Korringa-Kohn-Rostoker method in combination with the coherent potential approximation\cite{0034-4885-74-9-096501, SPRKKR}. To investigate the effect of the exchange-correlation functionals, we used several different implementations of the local density approximation and generalized gradient approximation. We found no significant dependence on the exchange-correlation functional for the quantitative values. Hence, we present only the values obtained from a full potential spin-polarized scalar relativistic calculation with the Perdew-Burke-Ernzerhof functional\cite{PhysRevLett.77.3865} for exchange correlations. The magnetic exchange integrals were calculated within the LKAG-formalism \cite{Liechtenstein198765}.
The $s$, $p$, $d$ and $f$ orbitals were included in the basis set. For accuracy of the exchange integral calculations, 20000 k-points were generated in the irreducible wedge of the Brillouin zone.

\subsection{1D Magnon diffusion model}
The proposed model is constructed for investigating if a non-homogeneous distribution of exchange interaction between free and localized spins ($J^{sd}$) can describe the delayed demagnetization of Ni relative to Fe in Py and Py-Cu. Here we assume an identical magnetic moment and inter-atomic exchange interaction between all atoms, i.e. the only property distinguishing Fe and Ni atoms is $J^{sd}$. In this model the $J^{sd}$ exchange interaction will preferentially generate magnons in the Fe sublattice that will diffuse and demagnetize Ni, resulting in a delayed Ni demagnetization relative to Fe.
First we will show that the magnon amplitude generated by this spatially dependent exchange interaction is described by $u(R_j,k_0)=I(R_j)^{1/2}\cdot e^{i(k_0\cdot R_j)}$, where $R_j$ is the atomic position and $k_0$ is the momentum transferred from the electronic to the spin system.
 If the amplitude $I(R_j)$ would be constant, then this describes a state with a well defined momentum $k_0$. However, when $I(R_j)$ is spatially varying, then $u(R_j,k_0)$ can be described as a superposition of states with a distribution of momentum ranging through the whole Brillouin zone. Importantly, even though $u(R_j,k_0)$ is described by a superposition of all possible momentum, the main intensity contribution is at $k_0$ (e.g. $\sim 90\%$ for parameters used in the optimal fit for Py). This means that the momentum will be relatively well defined and hence the distribution of $u(R_j,k_0)$, with regards to $k_0$, is given by a Bose-Einstein (BE) distribution (including a factor that incorporates the effect of the short time scales involved). The initial state is at this point fully known and described by the spatial distribution of $J^{sd}$, giving $I(R_j)$, and the BE distribution giving $k_0$. 
The time evolution of $u(R_j,k_0)$ will be obtained by evolving each momentum component separately according to the dispersion relation for magnons, and be subjected to a Gilbert damping. The time evolution will therefore be defined by the exchange stiffness and the Gilbert damping parameter. The spatial distribution of $u(R_j,k_0, t_n)$ will now  describe how the Fe sublattice and Ni sublattice are affected at time $t_n$. The total demagnetization at time $t_n$ is described by the sum of all excitations $\abs{u(R_j,k_0, t_n)}^2$ generated at different times and with different $k_0$, e.g. excitations at different times or with different $k_0$ are assumed to be incoherent. The number of excitations at each specific time is described by an exponentially decaying rate which is obtained by fitting this model to experimental data.

The spatial distribution of the magnon amplitude is determined by the Hamiltonian $H=H_0+H_1$, where $H_0=-2\sum_{i<j} J_{ij} \bold S_i \cdot \bold S_j$ is the exchange interaction between atomic sites i and j, and $H_1=-2 \sum_{j} J_{j} \bold s \cdot \bold S_j$ is the exchange interaction between free ($\bold s$) and localized ($\bold S_j$) spins. $H_1$ is the only term that provides a source for angular momentum to the localized moments and will be treated separately from $H_0$. The magnitude of the local magnetic moment ($\abs{\bold S_j}$) will be treated as equal for all lattice sites.
Following Ref.\ \cite{PhysRev.105.1439}, but including a site dependent exchange interaction, $H_1$ can be expressed as $H_1(k_0)= \frac{-2}{N}  \bold s \sum_j J_j^{sd}e^{ik_0\cdot R_j} \bold S_j$, where $N$ is the number of atoms, $R_j$ is the position of atom $j$ and $J_j^{sd}$ is the atomic s-d exchange interaction. The site dependent phase factor follows from the Block wave description of free electrons, which is here assumed to be valid in alloys and $k_0$ is the momentum transfer from the scattered electron to the the magnon system. Expanding $\bold s \cdot \bold S_j$ and using $S_x= \frac{S^++S^-}{2}$, $S_y= \frac{S^+-S^-}{2i}$, where $S^+$ and $S^-$ are creation and annihilation operators, we have $H_1(k_0)= \frac{-1}{N}  \sum_j J_j^{sd}e^{ik_0\cdot R_j} (s^-S_j^++s^+S_j^-+2s^zS^z_j)$. 
Even though the third term provide a momentum relaxation path for magnons, it does not involve angular momentum transfer between the free and localized spins\cite{PhysRev.105.1439} and will hence be ignored (momentum relaxation is partly accounted for by using a Bose-Einstein distribution described below).
Using the Holstein and Primakoff transformation and assuming small spin wave amplitudes, we have $S_j^+\approx(2S)^{1/2}a_j^-$, $S_j^-\approx(2S)^{1/2}a_j^+$, where $a_j^+$ and  $a_j^-$ are the magnon creation and annihilation operators. We therefore have $H_1(k_0)= \frac{-(2S)^{1/2}}{N}  \sum_j J_j^{sd}e^{ik_0\cdot R_j} (s^-a_j^-+s^+a_j^+)$. Following Ref.\ \cite{PhysRevB.77.144416}, the wave function can be expressed in the basis of Glauber coherent states corresponding to eigenstates of the annihilation operator $a_j^-\ket U=u(R_j)\ket U$, where $u(R_j)$ is the coherent amplitude of the annihilation operator at site j for the magnon state $\ket U$. The equation of motion is given by  $i\hbar\frac{da_j^-}{dt}=[a_j^-, H_1(k_0)]$, and since $[a_m^-,a_n^-]=0$ and $[a_m^-,a_n^+]=\delta_{mn}$, we have $\frac{da_j^-}{dt}=\frac{i(2S)^{1/2}s^+}{\hbar N} \cdot J_j^{sd}e^{ik_0\cdot R_j}$, giving $a_j^-=C(t)J_j^{sd}e^{ik_0\cdot R_j}$. The prefactor is site independent and given by $C(t)=\frac{i(2S)^{1/2}s^+}{\hbar N}\cdot t$. The site dependent coherent amplitude is then given by $u(R_j,k_0)=\bra U a_j^- \ket U=C(t) J_j^{sd}e^{ik_0\cdot R_j}$. Due to the exclusion of the $H_0$ term there is no dispersion in this description of the coherent amplitude, rendering it valid only on short timescales. However, due to the broad and random nature of the electronic excitations created by the laser, the coherence time is expected to be very short and we can therefore treat the excitations at different times as incoherent and independent from each other.

The dispersion is simulated using the above derived functional form of the space dependent wave function in a 1D lattice, $u(R_j,k_0)=I(R_j)^{1/2}\cdot e^{i(k_0\cdot R_j)}$, where $I(R_j)$ is real valued (a complex value implies a constant phase factor which does not influence the dynamics) and corresponds to the spatial magnon probability ($\lvert u(R_j,k_0) \rvert ^2=I(R_j)$, $\sum_j I(R_j)=1$). 
$I(R_j)$ is proportional to $(J^{sd}_j)^2$ and is hence determined by the random distribution of atomic species.  In k-space this results in a random distribution of magnon momentum that is strongly peaked at $k_0$.
The simulation is performed by creating a 1D lattice (1024 lattice sites) with a random position of Fe, Ni atoms, consistent with their respective concentrations $C_{Fe(Ni)}$.  The initial magnon distribution is given by the parameter $p$ which corresponds to the probability ratio of initially finding the magnon at an Fe atom relative to a Ni atom. $I(R_j)$ is then the spatial distribution of $p/(1+p)$ (for Fe) and $1/(1+p)$ (for Ni), normalized to unity when summed over the lattice, e.g. for all Fe and Ni concentrations, $p=1$ corresponds to a homogenous distribution of magnons in the lattice. The model cannot  account for the Cu atoms that have vanishing magnetic moment, hence Py and PyCu are treated equally i.e. Cu atoms are disregarded in the simulations. The lattice spacing corresponds to the nearest neighbor distance, where $L=2.475$ \AA\ has been used for all samples (bcc Fe, fcc Ni, fcc Py and fcc PyCu).

The dispersion is now achieved by evolving the phase of each Fourier component of $u(R_j,k_0)$ that are described by [$A(k,k_0)$, $P(k,k_0)$], where $A(k,k_0)$ is the k-dependent magnon amplitude and $P(k,k_0)$ is the phase.
The time dependent phase is $P(k,k_0,t)=\omega (k) \cdot t + P (k,k_0)$, where $\omega (k)=2D(1-cos(k\cdot L))/(\hbar L^2)$ is the magnon dispersion relation and D is the spin wave stiffness.
 The k-dependent Gilbert damping is included by a time dependent amplitude $A(k,k_0,t)=A(k,k_0)\cdot [1-A_{end}(1- e^{-\frac{t}{\tau_{G}(k)}})]^{1/2}$, where $\frac{1}{\tau_{G}(k)}=2\alpha_G\cdot \omega (k)$, $\alpha_G$ is the Gilbert damping parameter and $A_{end}$ is the fraction of magnons that has damped at long timescales. At time $t_n$, [$A(k,k_0,t_n)$, $P(k,k_0,t_n)$] is transformed back to real space, giving the spatial magnon distribution $S(R_j,k_0,t_n)=\abs {u(R_j,k_0,t_n)}^2$ of $k_0$ magnons at time $t_n$.
 The distribution of magnon momentum, $N(k_0)$, at ultrafast timescales has been approximated by $N(k_0)=N_{Norm}^{-1}\cdot \sqrt{\omega} \cdot BE(\omega(k_0), 700K)\cdot (1-e^{-t_{dis}(\frac{1}{\tau_{G}(k_0)}+\frac{1}{\tau_{L}(k_0)})})$, following Ref.\ \cite{PhysRevB.92.180412}. Here $N_{Norm}$ ensures that $\sum_{k_0}N(k_0)=1$, $\sqrt{\omega}$ corresponds to the 3D magnon density of states and $BE(\omega(k_0), 700K)$ is the Bose-Einstein distribution at 700 K. 
 Here we choose a constant value of $t_{dis}$ to provide a magnon distribution in the time interval of interest, where we use  $t_{dis}=400$ fs for investigating the remagnetization and $t_{dis}=100$ fs for the study of Ni demagnetization delay in Py and PyCu. Note that a fixed value of $t_{dis}$ is necessary for computational speed needed when using this model in a fitting procedure.
 The lifetime of magnons due to Landau damping, $\tau_L(k_0)$, is estimated from Ref.\ \cite{PhysRevLett.109.087203} using the relation $\tau_L(k_0)=\frac{2\hbar }{\Gamma(k_0)}$, where $\Gamma(k_0)$ is the lifetime broadening of high k-magnons. Here an analytical fit to the experimental data\cite{PhysRevLett.109.087203} has been used, given by 
$\Gamma(k_0)=0.131\cdot (k_0-k_L)$ eV, where $k_L=0.296$ \AA$^{-1}$ corresponds to the smallest value of $k_0$ that is susceptible to Landau scattering.
The total spatial magnon distribution is obtained by summing all the magnons with different momentum $k_0$, weighted by the distribution $N(k_0)$, i.e. $S(R_j,t_n)=\sum_{k_0} S(R_j,k_0,t_n) \cdot N(k_0)$. In all simulations we have used 80 $k_0$-points distributed evenly between $-\pi /L<k_0<\pi /L$.
From $S(R_j,t_n)$ one extracts $S_{Fe}(t_n)=\sum_{j=Fe}S(x,t_n)$ and $S_{Ni}(t_n)=\sum_{j=Ni}S(x,t_n)$, which are the probabilities to find the magnon in the Ni and Fe sublattice, respectively. Note that for $t_n>0$,  $S_{Fe}(t_n) + S_{Ni}(t_n)<1$ due to damping. So far $S_{Fe(Ni)}(t_n)$ only describes the time evolution of magnons that were generated at $t=0$. However, the demagnetization of a specific sublattice at time $t_n$  also needs to include magnons that were generated between $t_n$ and $t=0$. Due to the identical distribution and dynamics of magnons generated at different times, the sublattice magnon occupation at time $t_n$ of a magnon that was generated at time $t_0$, is given by $S_{Fe(Ni)}(t_n-t_0)$. Therefore, the total $S^{tot}_{Fe(Ni)}(t_n)$ sublattice occupation is given by the weighted sum over all $t_0<t_n$, i.e.  $S^{tot}_{Fe(Ni)}(t_n)=\int_0^{t_n}S_{Fe(Ni)}(t_n-t_0)\cdot R(t_0)dt_0$
The weighting factor, $R(t_0)dt_0$, is the number of magnons generated at time $t_0$ (normalized) and hence $R(t)$ is the normalized magnon generation rate.
The rate is given by $R(t)=[\frac{M_{end}}{\tau_{dem}}e^{(-t/\tau_{dem})}]*\gamma(t)$, where $\gamma(t)$ is a gaussian, convoluted due to the temporal width of the pump-pulse, $\tau_{dem}$ is the demagnetization time, and $M_{end}$ is the normalized amount of demagnetization. 
The FWHM of $\gamma$ was set to 40 fs for Fe, Ni, Py and 100 fs for PyCu. The normalized asymmetries $A_{Fe(Ni)}$, that are fitted to the experimental data, are then given by $A_{Fe(Ni)}(t)=1-(1\pm w)\cdot \frac{S^{tot}_{Fe(Ni)}(t)}{C_{Fe(Ni)}}$, where $w$ is a small correction for the mismatch in asymmetry between Fe and Ni at longer timescales (without this correction factor the model does not support a crossing between Fe and Ni demagnetization curves which is apparent in Fig.\ \ref{Demag_Py_Cu}a), $+w$ is used for the Ni asymmetry and $-w$ for Fe. 
For Py and PyCu, $w$ was set to 0.02 and 0.03, respectively.

%-----------------------------------------------------------------------------------------------

%\bibliography{Refs_CuPy}


\end{document}